\documentclass[10pt,twocolumn,english,preprint,prl,amsfonts,amssymb,byrevtex,tightenlines,nobalancelastpage]{revtex4}
\usepackage[T1]{fontenc}
\usepackage[latin9]{inputenc}
\usepackage{graphicx}

\makeatletter
\@ifundefined{textcolor}{}
{%
 \definecolor{BLACK}{gray}{0}
 \definecolor{WHITE}{gray}{1}
 \definecolor{RED}{rgb}{1,0,0}
 \definecolor{GREEN}{rgb}{0,1,0}
 \definecolor{BLUE}{rgb}{0,0,1}
 \definecolor{CYAN}{cmyk}{1,0,0,0}
 \definecolor{MAGENTA}{cmyk}{0,1,0,0}
 \definecolor{YELLOW}{cmyk}{0,0,1,0}
}

\providecommand{\U}[1]{\protect\rule{.1in}{.1in}}

\makeatother

\usepackage{babel}
\begin{document}

\title{Electrostatic Tuning of the Properties of Disordered Indium Oxide
Films near the Superconductor-Insulator Transition }

\author{Yeonbae Lee,$^{1}$ Aviad Frydman,$^{2}$ Tianran Chen,$^{1}$ Brian
Skinner,$^{1}$ and A. M. Goldman$^{1}$}

\affiliation{$^{1}$School of Physics and Astronomy, University of Minnesota,
116 Church St. SE, Minneapolis, MN 55455, USA \linebreak{}
$^{2}$Department of Physics, Bar Ilan University, Ramat Gan 52900,
Israel}
\begin{abstract}
The evolution with carrier concentration of the electrical properties
of amorphous indium oxide (InO$_{x}$) thin films has been studied
using electronic double layer transistor configurations. Carrier variations
of up to $7\times10^{14\,}carriers/cm^{2}$ were achieved using an
ionic liquid as a gate dielectric. The superconductor-insulator transition
was traversed and the magnitude and position of the large magnetoresistance
peak found in the insulating regime were modified. The systematic
variation of the magnetoresistance peak with charge concentration
was found to be qualitatively consistent with a simulation based on
a model involving granularity. 
\end{abstract}
\maketitle
The competition between disorder and superconductivity has long been
the subject of theoretical and experimental study\cite{Goldman,Steiner,Fisher,Imry,Trivedi,Dubi}.
By varying the nominal disorder a superconductor-insulator transition
(SIT), known as disorder-driven SIT, is found in materials such as
amorphous Bi \textit{(a-}Bi)\cite{Haviland}, TiNx\cite{Baturina},
and InOx\cite{Shahar and Zvi,Sambandamurthy}. Such transitions in
disordered thin films are believed to be quantum phase transitions.
Among the materials, amorphous InO$_{x}$ (which we will refer to
as InO) is of great interest because of the novel feature of a giant
magnetoresistance (MR) peak at low temperatures\cite{Paalanen,Sambandamurthy,Gantmakher,Shammass}.
Moreover direct evidence of Cooper pairs, both above the transition
temperature, and in the insulating regime has been reported\cite{Sacepe,Sherman}.
However, the study of the disorder-driven SIT in InO, mainly achieved
by varying its oxygen concentration during growth and heat treatment
afterward, can introduce unavoidable complexities to the system including
the conflation of variations of carrier concentration and levels of
disorder\cite{Gantmakher,Ovadyahu}. 

In this Letter, we overcome these limitations by using a different
approach. Namely, we tune the carrier concentration by adapting a
field effect transistor (FET) configuration, and we use this method
to induce a SIT within a given sample without altering its disorder
(In earlier work, Parendo \textit{et al}\cite{Parendo} reported the
study of the electrostatically tuned SIT of \textit{a-}Bi.). Benefiting
from the FET configuration, we present a comprehensive study of the
SIT driven by carrier modulation. This has revealed many interesting
phenomena such as variable range hopping in the insulating regime
and the broadening of superconducting fluctuations near the SIT. This
new approach also enables us to produce and then continuously change
the size and the position of MR peak in the insulating regime. We
believe that our findings shed light on the origin of the MR peak
and the possibility of localized superconducting islands within the
insulating regime.

\begin{figure}
\includegraphics[scale=0.95]{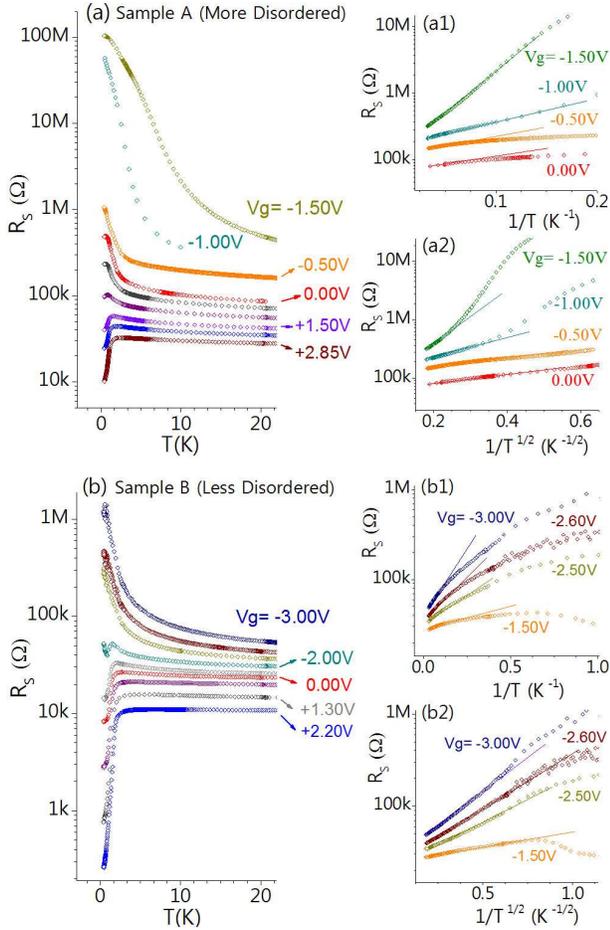}\caption{(Color online) The temperature dependence of $R_{s}$ at various values
of $V_{g}$ of samples A (a) and B (b). The onset of superconductivity
(the initial downward dip in the $R_{s}$ vs. $T$ curve) occurs at
$V_{g}=+1.00V$ for sample A, and at $V_{g}=-2.00V$ for sample B
respectively. On the right hand panels, both $R_{s}$ vs. 1/$T$ and
1/$T^{1/2}$ are plotted for sample A (a1-a2) and for sample B (b1-b2).
Note the cross-over from Arrhenius to ES VRH at $V_{g}=-0.50V$ shown
in (a1-a2), whereas (b1-b2) shows ES VRH over all ranges of $V_{g}$.}
\end{figure}

Amorphous (meaning non-crysal within our context) indium oxide (InOx)
films were e-gun evaporated using 99.999\% pure In$_{2}$O$_{3}$
pallets. The films were grown on Si wafers (100) through a shadow
mask defining a Hall bar geometry. The base pressure in the deposition
chamber was $2\times10^{-7}$mbar. Pure $\mbox{O}_{2}$ gas with partial
pressure $4\times10^{-6}$mbar was bled into the chamber during the
evaporation in order to achieve samples with sheet resistances of
about $10\, k\Omega$ at $T=4K$, close to the SIT. Sample thicknesses
ranged from 10 nm to 6 nm. The 6 nm samples were all insulating as
grown and are the central focus of this work. Gold electrodes, 50
nm thick, were deposited on the InO films. An ionic liquid (DEME-TFSI)
was used as a gate dielectric with a Pt coil top gate to form an electric
double layer transistor (EDLT). A schematic of the top-gated EDLT
can be found in our previous work\cite{Yeonbae}. (Earlier, Misra,
McCarthy, and Hebard gated InO films with ILs but did not explore
their superconductivity\cite{Hebard}.) The sheet resistances $R_{s}$
of the films were determined employing a four-probe electrode configuration.
A $^{3}$He refrigerator with a superconducting magnet enabled us
to vary the temperature between 300K and 0.40K and the magnetic field
between 0T and 9T. Gate voltages $V_{g}$, ranging from +3V to -3V
were used to induce or deplete charge carriers, which are electrons.
The various values of $V_{g}$ were applied at a temperature of 240K
and were held constant throughout the subsequent cooling and measurement. 

The $T$ dependencies of $R_{s}$, of two InO samples at different
values of $V_{g}$ are plotted in Fig. 1. Both samples were grown
under the same conditions\textit{ i.e}., at the same oxygen pressure
and at the same time. However their initial resistances were adjusted
so as to be different, by annealing at 65$^{o}C$ under vacuum for
different periods of time. Both samples exhibited large changes in
$R_{s}$ upon gating. Unfortunately, Hall effect measurements of such
highly disordered systems were not possible, except in the case of
a few other samples (not shown) that had relatively low normal state
resistances. The process of gating was found to be reversible, with
minor hysteresis. This suggests that apart from the remote possibility
of a reversible chemical reaction, the gating process is electrostatic.
To determine and/or confirm the carrier modulation of the system,
we used an electrochemical technique known as the Chronocoulometry\cite{Daghero}.
Using this technique, we observed changes in the sheet (2D) carrier
density or charge transfer ($\Delta n_{sheet})$ of up to $7\times10^{14}\, carriers/cm^{2}$.
However we did not characterize the charge transfer at each gate voltage
for the films reported here. 

The accessible charge transfer was large enough for us to observe
the SIT of InO as shown in Figs. 1(a) and 1(b). The right hand panels
of Fig. 1 show $R_{s}$ vs. $T$ behavior of the insulating state;
both Arrhenius $R=R_{0}exp(T_{0}/T)$ and Efros-Shklovskii variable
range hopping (ES VRH) $R=R_{0}exp(T_{0}/T)^{1/2}$ were observed.
Comparing Figs. 1(a1) and 1(b1), the higher resistance sample shows
a wider range of Arrhenius activated transport, suggesting that it
has a hard gap in its single-particle density of states. A cross-over
from Arrhenius to ES hopping is observed as the film becomes less
resistive with increasing carrier concentration. The observed ES hopping
suggests the presence of a soft gap (Coulomb gap) in the density of
states near the Fermi level, which is due to the long range interactions
between electrons in the system\cite{Efros}. This has also been reported
in other systems\cite{Baturina,Gerber}. The transition from Arrhenius
to ES hopping could be interpreted as the evidence of enhanced electron-electron
interaction driven by the addition of carriers. Both samples A and
B exhibited SI transitions tuned by carrier modulation, which is clearer
in the case of sample B {[}Fig. 1(b){]}. 

\begin{figure}
\includegraphics[scale=0.3]{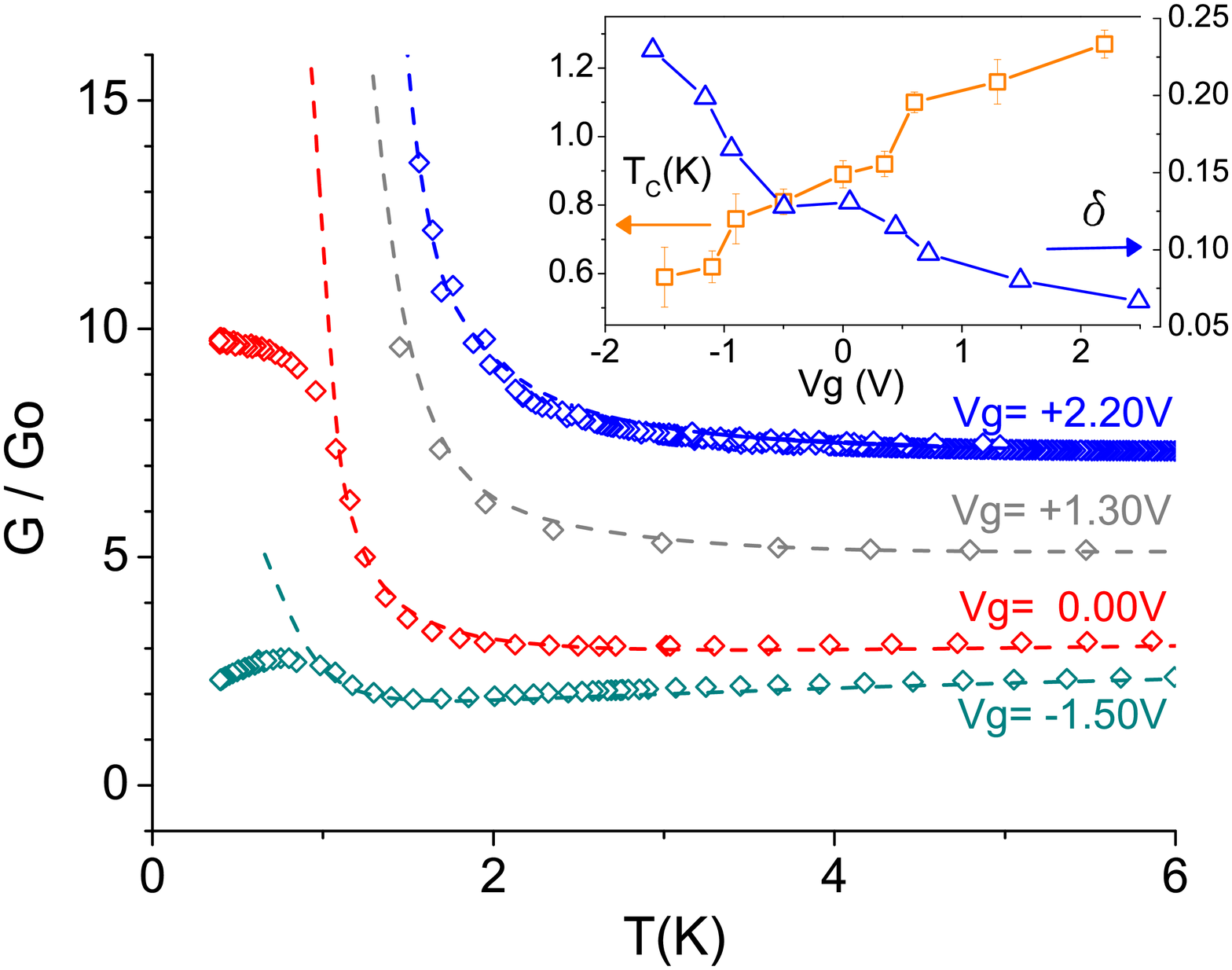}\caption{(Color online) Dimensionless conductance $G/G_{0}$ vs. $T$ for the
sample B in a superconducting state regime. The dashed lines are the
best fits to the data. The inset is a plot of $T_{c}$ (left) and
$\delta$ (right) vs. $V_{g}$.}
\end{figure}

When a system enters a superconducting state below a critical temperature
$T_{c}$, global phase-coherence is established with a nonzero amplitude
of the superconducting order parameter. Determining $T_{c}$ from
data, however, can be a non-trivial task because of the broadening
of the superconducting transition due to superconducting fluctuations
(SF) above $T_{c}$\cite{SF Book}. Thus, in order to estimate $T_{c}$,
we follow a recent theoretical model described in Ref. \cite{Varlamov}.
According to this model, superconducting fluctuations in 2D system
give rise to a temperature-dependent change in conductance $\Delta G$
given by $\Delta G=\Delta G^{AL+MT}+\Delta G^{DOS}+\Delta G^{WL+EE}$,
where $\Delta G^{AL+MT}$ is the combination of the Aslamazov-Larkin
(AL) and Maki-Thompson (MT) processes\cite{Thompson,Kawaguti}, $\Delta G^{DOS}$
accounts for the change in the single-particle density of states (DOS)
due to their involvement in fluctuation pairing\cite{Varlamov}, lastly
$\Delta G^{WL+EE}$ is the correction from the weak localization (WL)
theory including electron-electron (EE) interactions\cite{Altshuler}.
With estimated coherence length of 13\textasciitilde{}20 nm (Compare
this with the typical superconducting coherence length of 10\textasciitilde{}30
nm for InO\cite{Johansson}), the films with 6 nm thickness can be
treated as (quasi) 2D system enabling us the analysis of SF theory
in 2D model. To fit the SF fit to the data, we used the asymptotic
expression in Table 1 (regime I, the Ginzburg-Landau regime) of Ref.
\cite{Varlamov}. Figure 2 is the result of the SF fit applied to
the data of sample B and the values of both $T_{c}$ and pair breaking
parameter $\delta$ are determined simultaneously from the best fit
as shown in the inset. This analysis allows us to determine $T_{c}$
and we observe a monotonic increase in $T_{c}$ as function of $V_{g}$.
This result implies a monotonic increase of the superconducting pairing
energy $\Delta$ as function of carrier density.

The MR measurements of these two samples at different gate voltages
are shown in Fig. 3. The most significant feature of the data for
sample A {[}Figs. 3(a1), 3(a2), and 3(a3){]} is the transition from
negative to positive MR followed by a downward slope of $R_{s}$ (the
MR peak) upon further increase of $H$, all taken at fixed $T$, over
the range from 0.5K to 1K. Also note the strong MR peak found in insulating
regimes of both samples. The position of the peak field, $H_{peak}$,
at which the maximum resistance occurs, monotonically increase as
function of $V_{g}$.

\begin{figure}
\includegraphics{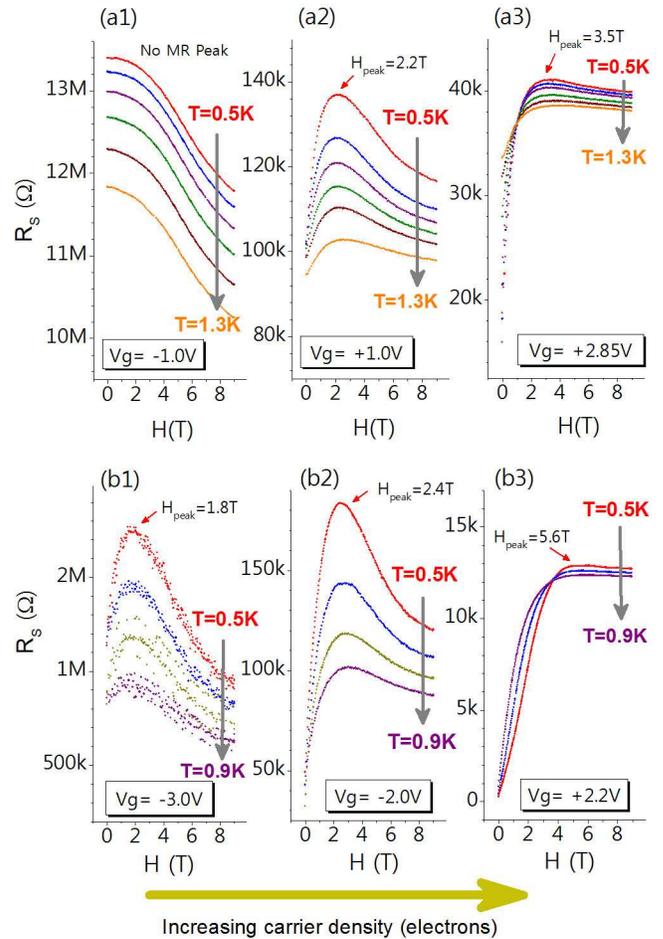}

\caption{(Color online) $R_{s}$ vs. $H$ at various values of $T$ and $V_{g}$
for sample A, labeled (a1-a3) and sample B, labeled (b1-b3). The density
of electrons increases from the left to the right. Each curve represents
a separate isotherm of $R_{s}$ vs. $H$, where $T$ ranges from 0.5\textasciitilde{}1.3K
for sample A (more disordered, top) and from 0.5\textasciitilde{}0.9K
for sample B (less disordered, bottom). }
\end{figure}

In a disorder-driven SIT, the spatial inhomogeneity of the pairing
energy $\Delta$ can be very important. In the insulating regime,
Ghosal et al.\cite{Trivedi} suggested that the system breaks up into
superconducting islands. Furthermore, in Ref. \cite{Dubi} such a
spatial variation of the order parameter amplitude has been demonstrated
using a negative U-Hubbard model. Near the SI transition, it has been
suggested\cite{Dubi-2,Meir} that the MR peaks in disordered systems
arise because magnetic fields affect the concentration and size of
superconducting islands, so that as these islands shrink with increasing
field, there is a transition from Cooper pair-dominated to single
electron-dominated transport. 

On the other hand, reduction of the superconducting pairing energy
$\Delta$ within islands can itself lead to a tradeoff between conduction
by Cooper pairs and conduction by unpaired electrons, and thus (potentially)
to a MR peak, even when the concentration and the size of the islands
are fixed. Recent theoretical works\cite{Mitchell,Tianran} studied
a model with fixed size and concentration of superconducting grains
(islands), and show how a MR peak deep in the insulating state can
arise as a result of the reduction of the superconducting gap $\Delta$
with increased $H$. This predicts that near the MR peak and at low
temperature the conduction should be described by ES VRH, as shown
in Figs. 1(a2) and 1(b2). Both approaches lead to an insulator in
which Cooper pairs with nonzero $\Delta$ are formed in the insulating
regime of the system and are responsible for the MR peak.

The shift of the MR peak to higher magnetic fields with increasing
carrier concentration, as shown in Fig. 3, can be explained qualitatively
within the context of the theory of Ref. \cite{Vinokur,Mitchell,Tianran}.
Increasing the carrier density presumably increases the density of
states at the Fermi level within the superconducting grains, thereby
driving up the zero-field superconducting gap, $\Delta_{0}$. A larger
$\Delta_{0}$ implies that a larger $H$ is required in order to reduce
$\Delta$ to the value of the grain charging energy $E_{c}$, so that
the MR peak shifts to higher $H$. In this way the transition from
negative MR {[}as in Fig. 3(a1){]} to a peak at an intermediate $H$
{[}Fig. 3(a2){]} to a peak at a larger $H$ {[}Fig. 3(a3){]} can be
understood. 

As an example, Fig. 4 shows values of the resistance of a simulated
2D array of regularly-spaced, monodispersed superconducting grains
as a function of $H$, calculated using the method described in Ref.
\cite{Tianran}. At small $\Delta_{0}/E_{c}$, the conductivity is
primarily due to hopping of unpaired electrons, and there is a monotonic
negative MR {[}as seen, for example, in Fig. 3(a1){]}. At larger $\Delta_{0}/E_{c}$,
which ostensibly corresponds to larger carrier density, the MR develops
a peak associated with a trade-off between conductivity by single
electrons and conductivity by Cooper pairs. This peak moves to larger
$H$ as $\Delta_{0}/E_{c}$ is increased {[}as in Fig. 3{]}. For the
simulation of Fig. 4 we have assumed a conventional BCS-like dependence
of $\Delta$ on the field $H$: $\Delta=\Delta_{0}\sqrt{1-(H/H_{c})^{2}}$.
In this way the data shown in Fig. 3 is consistent with the concept
of tuning the local superconducting gap by modulating the carrier
density. Such tuning of the gap by carrier density also suggests a
straightforward possible explanation of the transition from the insulating
state to the superconducting state with increasing carrier density.
Presumably, the transition may result from regions with zero $\Delta$
being driven to non-zero $\Delta$, thereby connecting superconducting
grains that would otherwise be disconnected. Unfortunately, the simulation
used to generate Fig. 4 cannot be used for a quantitative determination
of the relationship $\Delta(n)$, since this requires a knowledge
of the $H$-dependence of the gap as well as the relative localization
lengths $\xi_{1}$ and $\xi_{2}$ for unpaired and paired electron
hopping. We also caution that the simulation technique is applicable
only for the heavily-insulating limit, and in this sense our comparison
between Figs. 3 and 4 is only qualitative. It should also be noted
that within this simple model a strong MR peak develops only at a
relatively large ratio of the localization lengths, $\xi_{2}/\xi_{1}$.
A final caveat is the possibility that other models may give similar
results.

\begin{figure}
\includegraphics[scale=0.6]{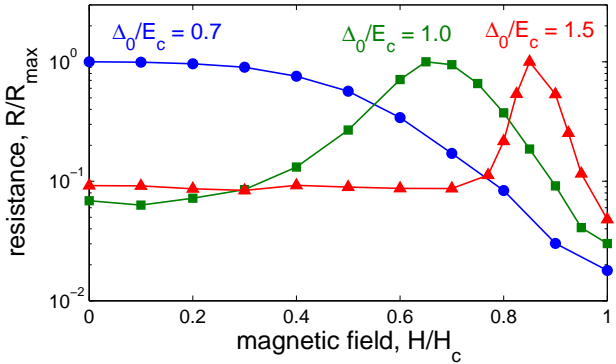}\caption{(Color online) Simulation of the resistance of a 2D array of identical
superconducting grains deep in the insulating state as a function
of $H$. Different curves are labeled with their corresponding values
of $\Delta_{0}/E_{c}$, and are normalized to their maximum resistance,
$R_{max}$. As $\Delta_{0}$ is increased, which presumably corresponds
to a larger carrier density, a MR peak develops that shifts to larger
magnetic field, in qualitative agreement with what is seen in Fig.
3. Here all curves correspond to a temperature such that $k_{B}T=0.1E_{c}$
and have localization lengths $\xi_{1}$ and $\xi_{2}$ for single-electron
and pair conductivity, respectively, satisfying $\xi_{2}/\xi_{1}=8$.}
\end{figure}

There is an apparent connection of granular-like superconducting behavior
in an amorphous film to sample morphology. Atomic force microscope
(AFM) images of these films shown in Fig. 5 (sample B) exhibited thickness
roughness which could be interpreted as an effective granularity.
In a recent work, ultra thin amorphous Bi films were patterned with
nano-scale honeycomb array of holes and MR peaks were observed in
several samples with non-uniform film thicknesses\cite{Nguyen}. On
the other hand the same experiment shows no MR Peaks in samples with
uniform film thicknesses\cite{Hollen}. It is also shown that the
thickness variations are primarily responsible for producing the localized
Cooper pairs in a Bi films\cite{Hollen02}, and the level of the disorder
can be tuned by the variation of the thickness in such systems\cite{Haviland}.
Furthermore the study of InO reported by \cite{Shammass} shows that
the different levels of MR anisotropy, which are indirect evidence
of the superconducting strength, have a strong relationship to different
values of thickness. These findings suggest that there is a close
relationship between effective granularity and the localization of
Cooper pairs resulting superconducting grains in an insulating matrix.
However, a granular morphology may not be essential for a system to
exhibit granular like behavior. The oxygen concentration may be the
key to determining an effective granularity in the amorphous InO system.
A recent experimental work\cite{Givan} showed that mesoscale spatial
variations in the oxygen concentration are present in nominally homogeneous
InO films. Such chemical inhomogeneity can induce local carrier density
fluctuations resulting in mesoscale inhomogeneity of the pairing energy
$\Delta$. It has also been demonstrated that, by STM measurement,
such inhomogeneity can be present even when the sample surface shows
high-level smoothness\cite{Sacepe}. We also like to add a cautious
caveat that our system might indeed be a granular system due to the
strong morphology fluctuation at the surface.

\begin{figure}
\includegraphics[scale=0.2]{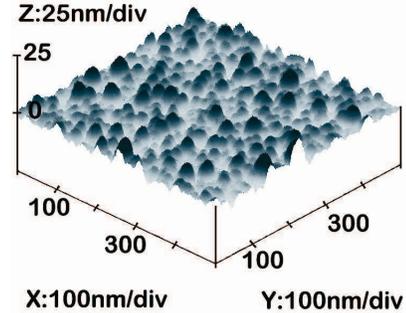}

\caption{(Color online) The AFM image for sample B revealing strong degree
of surface roughness. Similar results were observed for sample A.}
\end{figure}

In conclusion, we have observed the superconductor-insulator transition
along with the evolution of the MR peak in amorphous InO films by
varying their carrier density. A transition from a negative MR to
a strong MR peak followed by the suppression of the MR peak was observed
near the SIT by adding carriers. Strong MR peaks when the systems
are insulating support the presence of localized Cooper pairs in the
insulating regime and are qualitatively consistent with models based
on there being an effective superconducting granularity. Comparing
the data with a theoretical simulation we argue that larger carrier
density increases the paring energy $\Delta$, and therefore is responsible
for both the SIT and the monotonic increase of the MR peak position
to higher values of $H$. The EDLT configuration opens up the possibility
of carrier-driven SIT studies of many other disordered systems.

We would like to thank Zvi Ovadyahu, Andrei Varlamov, Nandini Trivedi,
Joe Mitchell, and Anirban Gangopadhyay for fruitful discussions and
advice. This work was supported by the NSF under Grant No. NSF/DMR-0854752,
by the NSF through the University of Minnesota MRSEC under Grant No.
NSF/DMR-0819885, and by the US-Israel Binational Science Foundation
under Grant No. 208299. Part of this work was carried out at the University
of Minnesota Characterization Facility, a member of the NSF-funded
Materials Research Facilities Network via the MRSEC program, and the
Nanofabrication Center which receives partial support from the NSF
through the NNIN program.

\end{document}